\documentclass{aa}  
\usepackage{graphicx}
\usepackage[varg]{txfonts}
\usepackage{hyperref}
\usepackage{natbib}
\usepackage{lscape}
\bibpunct{(}{)}{;}{a}{}{,}
\usepackage{epstopdf}
\begin{document}

\title{Abundances of disk and bulge giants from hi-res optical spectra\thanks{Based on observations made with the Nordic Optical Telescope (programs 51-018 and 53-002), operated by the Nordic Optical Telescope Scientific Association at the Observatorio del Roque de los Muchachos, La Palma, Spain, of the Instituto de Astrofisica de Canarias, and on spectral data retrieved from PolarBase at Observatoire Midi Pyrénées.}}
\subtitle{I. O, Mg, Ca, and Ti in the Solar neighborhood and Kepler field samples}
\author{H.~J\"onsson\inst{1,2,3} \and N.~Ryde\inst{1} \and T.~Nordlander\inst{4} \and A.~Pehlivan\inst{1,5} \and H.~Hartman\inst{1,5} \and P.~J\"onsson\inst{5} \and K.~Eriksson\inst{4}}
\institute{Lund Observatory, Department of Astronomy and Theoretical Physics, Lund University, Box 43, SE-221 00 Lund, Sweden\\ \email{henrikj@astro.lu.se}\and
Instituto de Astrofísica de Canarias (IAC), E-38205 La Laguna, Tenerife, Spain\and 
Universidad de La Laguna, Dpto. Astrofísica, E-38206 La Laguna, Tenerife, Spain\and
Department of Physics and Astronomy, Uppsala University, Box 516, SE-751 20 Uppsala, Sweden\and
Materials Science and Applied Mathematics, Malm\"o University, SE-205 06 Malm\"o, Sweden
}
	    
 \date{Submitted 2016; accepted 2016}

\abstract
   {The galactic bulge is an intriguing and significant part of our galaxy, but it is hard to observe, being both distant and covered by dust in the disk. Therefore there do not exist many high-resolution optical spectra of bulge stars with large wavelength coverage, whose determined abundances can be compared with nearby, similarly analyzed stellar samples.}
   {We aim to determine the, for chemical evolution models, so important alpha elements of a sample of bulge giants using high-resolution optical spectra with large wavelength coverage. The abundances found will be compared to similarly derived abundances from similar spectra of similar stars in the local thin and thick disks. In this first paper we focus on the Solar neighborhood reference sample.} 
   {We use spectral synthesis to derive the stellar parameters as well as the elemental abundances of both the local as well as the bulge samples of giants. Special care is taken to benchmark our method of determining stellar parameters against independent measurements of effective temperatures from angular diameter measurements and surface gravities from asteroseismology.}
   {In this first paper we present the method used to determine the stellar parameters as well as the elemental abundances, evaluate them, and present the results for our local disk sample of 291 giants.}
   {When comparing our determined spectroscopic temperatures to those derived from angular diameter measurements, we reproduce these with a systematic difference of +10 K and a standard deviation of 53 K. The spectroscopic gravities are reproducing the ones determined from asteroseismology with a systematic offset of +0.10 dex and a standard deviation of 0.12 dex. When it comes to the abundance trends, our sample of local disk giants is closely following that of other works analyzing solar neighborhood dwarfs, showing that the much brighter giant stars are as good abundance probes as the often used dwarfs.}

\maketitle

\section{Introduction} \label{sec:introduction}
How the galactic bulge formed and evolved is an intriguing question that is been given a lot of attention the last years \citep[see for example][for reviews]{2013pss5.book..271R,2015arXiv150307252G,2016arXiv160305485D}. There are three problems with observing the bulge; it is distant, so the stars are faint, it is obscured by the dust in the disk, so very little of the optical light emitted by the bulge stars reach us, and the crowding of stars is making it hard to place the spectrometer's slit to avoid contaminating light from another star. For the first two problems it helps to observe luminous giant stars, and indeed this has been the main method used, see for example \citet{2006A&A...457L...1Z,2007A&A...465..799L,2010A&A...513A..35A,2011A&A...530A..54G,2012ApJ...749..175J,2013ApJ...775L..27J,2014AJ....148...67J}.     

One fact that could cause problems when analysing spectra from giant stars, is the presence of molecules in their atmospheres. The molecules emerge since giant stars are generally cooler than the dwarf stars often used in spectroscopic investigations. A complication  is therefore the increased number of lines, mainly due to the rich molecular spectra. This might lead, particularly for the very coolest and most metal-rich stars, to the situation where the continuum close to lines of interest can not be identified even at very high spectral resolution. Since the abundance derived from a spectral line depends on the ratio of line to continuum opacities, large uncertainties will be introduced if the continuum can not be defined. A large density of lines might also form a  psuedo-continuum, which falsely might be taken as a true continuum, further complicating the analysis. The probability of spectral lines being blended, in the worst case by unknown lines or lines with uncertain line data, is increased by the molecular lines. The spectral lines of choice to use in an abundance analysis, which are as unblended as possible and not too deep, might therefore be rare. As this problem is worse the cooler a star is, many works, including the bulge-articles mentioned in the first paragraph (and this project), have restricted themselves to the moderately cool K-giants or red clump stars.

Furthermore, for sun-like stars, the abundance could be determined differentially by simply comparing the strength of the spectral line in question in the stellar spectrum to that of a solar spectrum. Such an approach for an analysis of giants is more complicated, and might introduce large systematic uncertainties: if a solar spectrum is used in the comparison, in many cases unknown blending lines show up in the giant star spectra, leading to an overestimation of the abundance of the element in the giant star, and if instead a spectrum of a giant star is used as a comparison, the zero-point of the derived abundances are very uncertain, because they rely on the accuracy of the abundances and stellar parameters of the comparison star, as well as the completeness of the linelist. 

To deal with these problems when giant stars are used in abundance works investigating for example the bulge, a good approach is to make a strictly differential comparison between the abundances found in the bulge to that of the more known stellar population of the local disk. For example, \citet{2010A&A...513A..35A}, found elemental abundances from their 25 bulge giants to be similar to that of the thick disk in their 55 similar giants from the Solar neighborhood, and \citet{2013A&A...549A.147B}, using microlensed dwarfs, found that the `knee' in their [$\alpha$/Fe] vs. [Fe/H]-plots likely was at slightly higher metallicities in their 58 bulge dwarfs as compared to their local disk sample of 714 stars \citep{2014A&A...562A..71B}.

In this paper we present a compilation of a solar neighborhood sample of 291 local disk giants for which we have determined the stellar parameters and the abundances of the $\alpha$ elements oxygen, magnesium, calcium, and titanium. We demonstrate that we can determine the elemental abundances of giants with similar precision and accuracy as for dwarfs. In Paper II of this series, we will use this solar neighborhood sample to differentially compare elemental abundances of giants in the bulge to abundances in the solar neighborhood. This will be possible since we will determine the stellar parameters and $\alpha$ abundances of the bulge sample in the same way.
 
The homogeneously analyzed local disk sample presented here, might also be useful in other aspects: since it includes stars in the Kepler field, the determined parameters might be used to revise the stellar properties for these stars \citep{2014ApJS..211....2H}, or the bright stars of the sample might be used as a basis for selecting and analyzing stars using smaller telescopes and/or less sensitive instruments, as for example the MIR spectrometer TEXES \citep{2002PASP..114..153L}. To our knowledge, the sample presented here is the largest spectroscopically analyzed sample of metal-rich giants using high-resolution optical spectra with large wavelength coverage.

\section{Observations} \label{sec:observations}
\subsection{The bulge sample}
As mentioned in Section \ref{sec:introduction}, the main purpose of this project is to analyze high-resolution optical spectra of bulge giants, but the analysis of the actual bulge spectra will be presented in Paper II of the series, while this paper concentrates on the analysis of a local disk comparison sample. The bulge sample consists of 46 K-giants observed with FLAMES/UVES at VLT, and consists of 35 spectra that have been previously analyzed in \citet{2006A&A...457L...1Z}, \citet{2007A&A...465..799L}, \citet{2010A&A...509A..20R}, \citet{2013A&A...559A...5B}, and \citet{2016A&A...586A...1V} as well as 11 never before analyzed spectra from a new bulge field even closer to the galactic center with $(l,b)=(1.25,-2.65)$. More details on this sample will be given in Paper II.

\subsection{The Solar neighborhood sample}
In this paper, Paper I in the series, we present the analysis of 291 giants in the Solar neighborhood: 150 were observed by us using the spectrometer FIES \citep{2014AN....335...41T} mounted on the Nordic Optical Telescope (NOT) under program 51-018 (May-June 2015), 63 more stars were observed by us using FIES/NOT under program 53-002 (June 2016), 41 spectra were taken from \citet{2012A&A...543A.160T} (in turn from FIES/NOT), 18 spectra were downloaded from the FIES-archive, and 19 spectra were downloaded from the NARVAL and ESPaDOnS spectral archive PolarBase \citep{2014PASP..126..469P}.

The FIES-spectra have $R\sim67000$ and the PolarBase-spectra have $R\sim65000$. They both cover the entire optical part of the spectrum, but only the region between 5800~\AA ~and 6800~\AA~is used in the analysis, to be consistent with the analysis of the bulge spectra in Paper II.

Most of the stars observed are very bright, and have typical observing times of the order of minutes. The 213 stars observed by us using FIES were observed using the `exp-count'-feature, aborting the exposure when a specified CCD count-level have been reached. Therefore, a vast majority of these spectra have a S/N of 80-120, see Table \ref{tab:basicdata} (Online material). The spectra downloaded from PolarBase generally have similar S/N, the FIES-archive spectra generally have slightly higher S/N, while the spectra from \citet{2012A&A...543A.160T} have lower S/N, around 30-50. All S/N as measured by the IDL-routine \texttt{der\textunderscore snr.pro}\footnote{See \href{http://www.stecf.org/software/ASTROsoft/DER\textunderscore SNR}{http://www.stecf.org/software/ASTROsoft/DER\textunderscore SNR}} are listed in Table \ref{tab:basicdata} (Online material).

Our FIES-spectra were reduced using the standard FIES-pipeline, while the spectra from \citet{2012A&A...543A.160T} and PolarBase were already reduced and ready to analyze. All spectra have been roughly normalized using the IRAF task \texttt{continuum}. In the analysis-step a more careful continuum-normalization is made for every wavelength window analyzed (see Section \ref{sec:analysis}).

\begin{figure*}[htp]
\centering
\includegraphics[width=180mm]{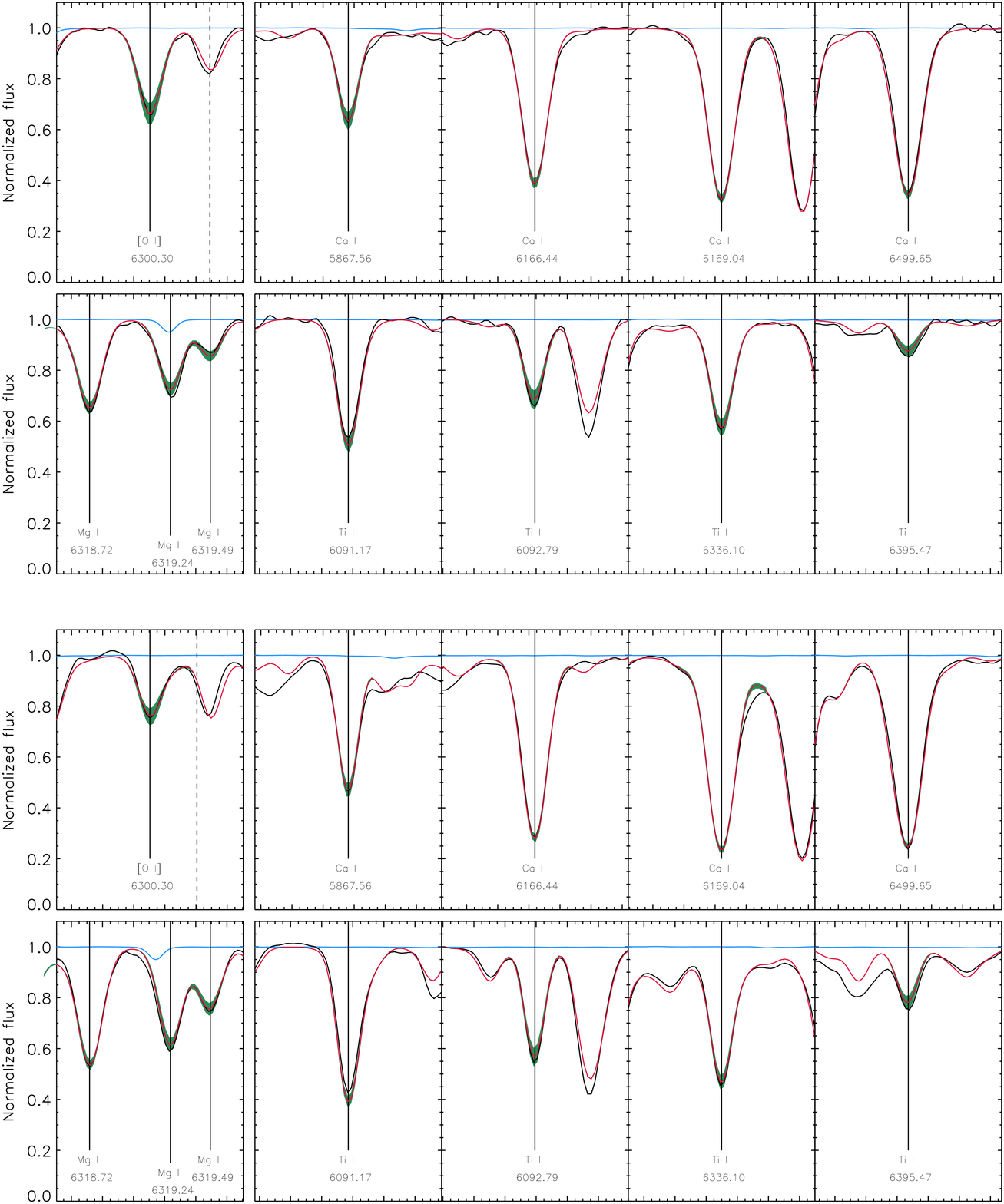}
\caption{The stellar lines used for abundance determinations in the analyzed FIES-spectra of Arcturus and $\mu$Leo, in the top and bottom two rows respectively. The stellar spectrum is shown in black, the best fitting synthetic spectrum  in red, and $\pm0.1$ dex of the element in question in green. The telluric spectrum from the Arcturus atlas of \citet{2000vnia.book.....H} is shown in blue and the position of the bright 6300 \AA~sky emission line is shown using a dashed line. Stellar lines possibly affected by telluric absorption lines and/or sky emission lines were avoided in the analysis. All panels show 1.2 \AA~of spectra surrounding the line in question, i.e. the larger tickmarks mark steps of 0.2 \AA.}
\label{fig:spectra}
\end{figure*}

No sky-subtraction and/or removal of telluric lines were attempted, instead regions of the spectra influenced by telluric absorption lines and bright sky emission lines were avoided: as can be seen in Figure \ref{fig:spectra}, telluric lines are affecting regions around the stellar oxygen and magnesium lines, while only the stellar oxygen line is affected by a possible sky emission line. The telluric lines and sky-line affecting regions around the widely used 6300 \AA oxygen line are important sources of uncertainties in the derived oxygen abundances in all works using this line.

The basic data for the analyzed disk giant stars are listed in Table \ref{tab:basicdata} (Online material).
 
\section{Analysis} \label{sec:analysis}
The 291 local disk spectra analyzed in this paper, as well as the 46 bulge spectra analyzed in Paper II were analyzed in the exact same way to ensure a strictly differential comparison.

We used the software \texttt{Spectroscopy Made Easy}, SME \citep{1996A&AS..118..595V}. SME simultaneously fits stellar parameters and/or abundances by fitting calculated synthetic spectra to an observed spectrum using $\chi^2$-minimization. By selecting regions with spectral lines of interest and points which SME should treat as continuum points, specific lines can be chosen as basis of the analysis. Special care was taken to avoid fitting spectral regions affected by telluric lines, particularly a problem for the oxygen and magnesium lines used. 

We have used spherical symmetric, [$\alpha$/Fe]-enhanced, LTE MARCS-models in the analysis. Within the Gaia-ESO collaboration \citep{2012Msngr.147...25G} SME has been modified to apply NLTE departure coefficients interpolated from the grid presented by \citet{2012MNRAS.427...50L}, which covers the stellar parameters and lines used in the analysis.

We determine all the stellar parameters ($T_\textrm{eff}$, $\log g$, [Fe/H], and  $\xi_\textrm{micro}$) simultaneously, using relatively weak, unblended $\ion{Fe}{i}$, $\ion{Fe}{ii}$, and $\ion{Ca}{i}$ lines and gravity-sensitive $\ion{Ca}{i}$-wings \citep{1988A&A...190..148E}. This means that the Ca-abundance is fitted simultaneously as the stellar parameters, but all other abundances are determined with fixed stellar parameters.  The atomic data of the spectral lines used are listed in Table \ref{tab:linedata} (Online material).

\subsection{Line data}\label{sec:linedata}
All line-data used, in both determining the stellar parameters as well as the abundances, have been taken from version 5 of the Gaia-ESO line-list \citep{2015PhyS...90e4010H} with some exceptions: the wavelength-data of all $\ion{Fe}{ii}$-lines have been updated with values from \citet{2013ApJS..204....1N}, and the $\log gf$-values have been updated from the calculated values of \citet{RU} to the astrophysical values of \citet{2009A&A...497..611M}. The latter values produce better fitting synthetic spectra, and most importantly, produce spectroscopic surface gravities closer to the asteroseismic surface gravities for the Kepler-stars in our sample. Furthermore the $\log gf$-values of the three $\ion{Mg}{i}$-lines used, have been updated from the calculated values of \citet{1993JPhB...26.4409B} to values from Pehlivan et al. (in prep). There is an autoionizing $\ion{Ca}{i}$-line close to these $\ion{Mg}{i}$-lines producing a very wide (up to 5 \AA) and shallow depression in the spectra. We found that removing this line from the synthesis, and instead placing a local continuum around $\ion{Mg}{i}$-lines produced the tightest magnesium trend in our data. Therefore no atomic data for the autoionizing $\ion{Ca}{i}$-line is used. Since we use the 6300.3083~\AA~[$\ion{O}{i}$]-line to determine oxygen abundance, the atomic data of the blending $\ion{Ni}{i}$-line at 6300.3419~\AA~is also of importance: here we have used the experimental $\log gf=-2.11$ from \citet{2003ApJ...584L.107J}. In the analysis of the different stars, a solar Ni-abundance scaled with the iron abundance of that particular star was used.

The line data of the lines used is listed in Table \ref{tab:linedata} (Online material). The final line list consists of 47 lines, of which 39 are used to determine the stellar parameters and the Ca-abundance. All lines are situated in the wavelength range between 5800 Å to 6800 Å to match the usable range of the bulge spectra analyzed in Paper II, thereby enabling a strictly differential comparison. 

Regarding the broadening of spectral lines due to collisions with neutral hydrogen (van der Waals broadening) the data for all the listed lines are taken from \citet{BA-J} and \citet{BPM}, with some exceptions not available in those references: for the \ion{Fe}{i} line at 6793 Å values from \citet{K07} are used, for the \ion{Ca}{i} line at 5867 Å values from \citet{S} are used, for the 6300 Å [\ion{O}{i}]-line vales from \citet{WSG} are used, and for the three \ion{Mg}{i}-lines values from \citet{KP} are used.

To minimize the errors and uncertainties in the determined stellar parameters and abundances we have carefully checked all used lines for (known) blends in a grid of stellar atmospheres, using a similar method as was used in \citet{2011A&A...530A.144J}. Typically a massive wealth of TiO molecular lines makes most lines in in our wavelength range very hard to use for stars with temperatures below $\sim3900$ K.

\subsection{Random uncertainties in the stellar parameters}\label{sec:random}
\begin{figure*}[htp]
\centering
\includegraphics[width=180mm]{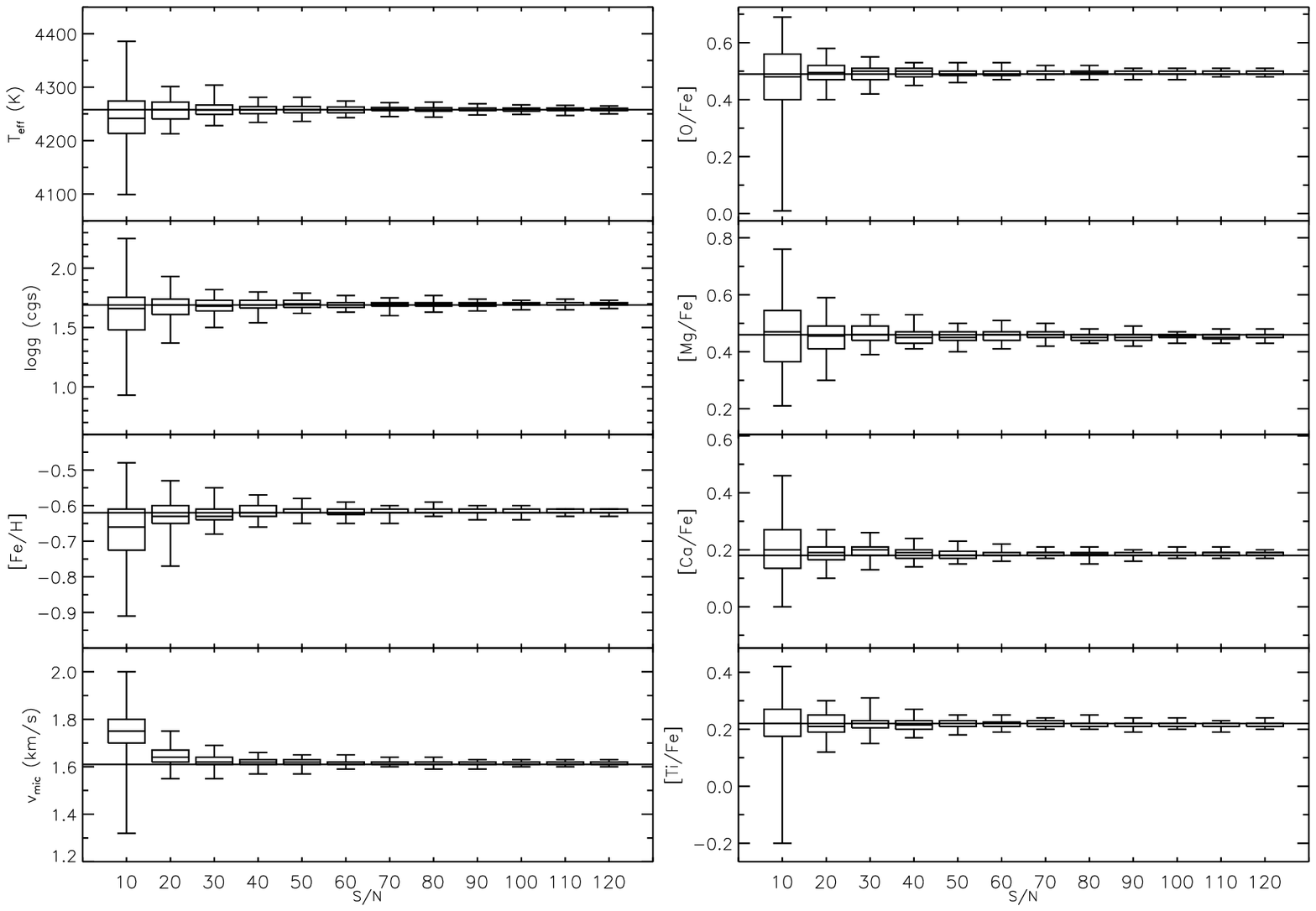}
\caption{Results from determining the stellar parameters and abundances for Arcturus spectra with different injected S/N. The horizontal line crossing each panel represents the value as determined from the original, high-S/N, atlas spectrum. The horizontal line in the boxes shows the median of the data, the lower and upper boundaries of the boxes show the lower and upper quartiles of the data, and the whiskers extend to the lowest and highest value of the data.}
\label{fig:snr}
\end{figure*}

Random uncertainties in our  method of determining the stellar parameters include both the freedom in setting the continuum and in the actual line fit. These obviously depend on the S/N of the observation in question. To test this, we have degraded the Arcturus spectrum of \citet{2000vnia.book.....H} to different S/N and determined the stellar parameters for those spectra. The \texttt{IDL}-routine \texttt{x\_addnoise}\footnote{\href{http://www.ucolick.org/~xavier/IDL/}{http://www.ucolick.org/~xavier/IDL/}} was used to inject noise and create 100 realizations each of Arcturus-spectra with S/N of 10 to 120 in steps of 10. The parameters for these spectra were then derived exactly as for the science spectra, and the results are shown in the box-plots in Figure \ref{fig:snr}.

From Figure \ref{fig:snr} one can deduce that for a S/N$\gtrsim50$ the uncertainties are of the same order, while they grow very large for S/N$\lesssim20$, meaning that integration times, at least for Arcturus-like stars, preferably should be adjusted to reach a S/N>20, but do not have to be so long that a much greater S/N than 50 is reached. However cooler and/or more metal-rich stars likely need higher S/N to resolve the more numerous lines, and also higher S/N is desired when determining, for example, the oxygen abundance that is based on a single line (see Table \ref{tab:linedata}, Online material). Of interest is also that it is mainly the `whiskers' of the plots that are expanding for lower S/N, the `boxes', holding 50\% of the data, is more similarly sized as the S/N is lowered. This means that, even for the lowest S/N-bins, we are determining reasonable stellar parameters for a majority of the spectra.  

Since all our program stars are bright, they have excellent S/N, in general around 100, and we can therefore, based on Figure \ref{fig:snr}, conclude that the random uncertainties stemming from the quality of the spectrum for an Arcturus-like red giant would be very small, of the order $\delta T_{\mathrm{eff}}=\pm10$~K, $\delta \log g=\pm0.05$, $\delta$[Fe/H]$=\pm0.02$, $\delta v_{\mathrm{mic}}=\pm0.02$ (standard deviation). Instead our uncertainties are dominated by the systematic uncertainties evaluated in the next section.

In Paper II, the uncertainties of the stellar parameters and abundances derived from the  observations of the very faint bulge stars will in many cases be dominated by the random uncertainties stemming from the S/N.

\subsection{Systematic uncertainties in the stellar parameters}\label{sec:systematic}
Systematic uncertainties in stellar parameters and abundances are very hard to assess: they stem from everything from uncertainties in the atomic data used, to simplifications in the stellar atmosphere models. To asses these systematic uncertainties of the stellar parameters, we  evaluate them in three steps by comparison with trusted benchmark values:
\begin{itemize}
  
\item In our sample, we have observed three of the giants of the Gaia-ESO benchmark stars \citep{2015A&A...582A..49H,2014A&A...564A.133J,2015A&A...582A..81J} with well determined parameters and abundances. Our results and the benchmark values are listed in Table \ref{tab:ges}.

\begin{table*}[ht]
\caption{The Gaia-ESO stellar parameter benchmark values \citep{2015A&A...582A..49H,2014A&A...564A.133J,2015A&A...582A..81J} are listed for the three giants in our sample. On the row below are the stellar values determined using our analysis for the corresponding star. In the case of Arcturus (HIP68673), we have two determinations: the first row shows our results using our FIES-spectrum, and the second row shows our results using the atlas of \citet{2000vnia.book.....H}. [Fe/H] is listed in the \citep{2009ARA&A..47..481A} scale, i.e. with A(Fe)$_{\odot}=7.50$}
\begin{tabular}{l l c c c c c c c c c c}
\hline
\hline
HIP & Name & $T_{\mathrm{eff}}$ & $\log g$ & [Fe/H] & $v_{\mathrm{mic}}$ & A(O) & A(Mg) & A(Ca) & A(Ti)\\ 

\hline
HIP37826   &  $\beta$~Gem    & $4858\pm60$ & $2.90\pm0.08$ & $0.08\pm0.16$  & $1.28\pm0.21$ &  ... & $7.56\pm0.07$  & $6.40\pm0.08$ & $4.96\pm0.07$ \\
           &                 &  4835       &  2.93         &  0.04          &  1.24         & 8.69 &  7.63          &  6.43         &  4.92 \\
           \hline
HIP48455   &  $\mu$~Leo      & $4474\pm60$ & $2.51\pm0.11$ & $0.20\pm0.15$  & $1.28\pm0.26$ &  ... & $8.11\pm0.11$  & $6.60\pm0.12$ & $5.22\pm0.10$ \\
           &                 &  4461       &  2.65         &  0.20          &  1.55         & 8.93 &  7.83          &  6.50         &  5.13 \\
           \hline
HIP69673   & $\alpha$Boo     & $4286\pm35$ & $1.64\pm0.09$ & $-0.57\pm0.08$ & $1.58\pm0.12$ & ...  & $7.49\pm0.09$  & $5.92\pm0.13$ & $4.59\pm0.08$ \\
           &                 &  4251       &  1.72         & $-0.60$        &  1.64         & 8.57 &  7.38          &  5.88         &  4.54 \\
           &                 &  4258       &  1.69         & $-0.62$        &  1.61         & 8.56 &  7.44          &  5.91         &  4.55 \\

\hline
\end{tabular}
\label{tab:ges}
\end{table*}

For these three stars, we find our parameters to be within the uncertainties of the benchmark values with one exception: our $\log g$ for $\mu$Leo is slightly higher. In general all our surface gravities are slightly higher than the three benchmark values, but for the other two stars our results are as stated within the uncertainties.

\item Furthermore we note that, except for Arcturus, $\beta$~Gem, and $\mu$~Leo, nine of our program stars have temperatures determined from angular diameter measurements \citep{2003AJ....126.2502M}, see Table \ref{tab:moz}.

\begin{table}[ht]
\caption{The effective temperatures as derived based on angular diameter measurements \citep{2003AJ....126.2502M} listed for the nine giants in our sample. Also listed are the stellar parameters determined using our analysis based on observations from a FIES, NARVAL, or ESpaDOnS spectrum for the corresponding star.}
\begin{tabular}{l l c c c c}
\hline
\hline
HIP & Name & $T_{\mathrm{eff, ref}}$ & $T_{\mathrm{eff}}$ & $\log g$ & [Fe/H]\\ 
\hline
     HIP9884 & $\alpha$~Ari   & 4493 $\pm$ 55 &   4464 & 2.27 &  -0.24 \\
    HIP46390 & $\alpha$~Hya   & 4060 $\pm$ 50 &   4095 & 1.56 &  -0.10 \\
    HIP54539 & $\psi$~UMa     & 4550 $\pm$ 56 &   4534 & 2.33 &  -0.10 \\
    HIP55219 & $\nu$~UMa      & 4091 $\pm$ 50 &   4133 & 1.65 &  -0.17 \\
    HIP72607 & $\beta$~UMi    & 3849 $\pm$ 47 &   3992 & 1.32 &  -0.23 \\
    HIP74666 & $\delta$~Boo   & 4850 $\pm$ 60 &   4861 & 2.63 &  -0.37 \\
    HIP77070 & $\alpha$~Ser   & 4558 $\pm$ 56 &   4540 & 2.61 &   0.16 \\
    HIP94376 & $\delta$~Dra   & 4851 $\pm$ 67 &   4807 & 2.71 &  -0.17 \\
   HIP102488 & $\epsilon$~Cyg & 4756 $\pm$ 59 &   4711 & 2.59 &  -0.18 \\
\hline
\end{tabular}
\label{tab:moz}
\end{table}

Our determined temperatures are within the uncertainties from the temperatures of \citet{2003AJ....126.2502M}, with one exception: $\beta$~UMi, where we derive a temperature 143~K higher than the reference value, showing the difficulties in determining the temperatures for the very coolest stars. All in all, we are able to derive the temperatures of these stars with a systematic offset of +10~K and a standard deviation of 53~K, very similar to the mean of the uncertainties of the measurements of \citet{2003AJ....126.2502M}: 56~K.

\item When it comes to the surface gravity, our sample includes 39 giants in the Kepler field with asteroseismic gravities \citep{2012A&A...543A.160T,2014ApJS..211....2H}: our determined gravities are deviating from the seismic values with a systematic offset of +0.10 dex and a standard deviation of 0.12 dex.
\end{itemize} 

To conclude, our determined effective temperatures seem very precise with a systematic shift of only +10~K, while we likely have a systematic shift of $+0.1$ dex in surface gravity. Regarding the metallicity, we only have three benchmark values, but our results are accurate with respect to those.

\subsection{Uncertainties in the determined abundances}\label{sec:uncertainties}
When it comes to the uncertainties in the determined abundances, we assess and estimate them below in three steps:
\begin{itemize}
    \item In the rightmost panels of Figure \ref{fig:snr}, we show the determined abundances in the S/N-injected spectra, using the stellar parameters as determined in the same S/N-injected spectrum (those that are shown in the leftmost panels). As expected, the uncertainties in the determined abundances follow the uncertainties in the stellar parameters, and show very large uncertainties for the lowest S/N-bin. However, just like for the stellar parameters, the box, holding 50\% of the values is still reasonably small, meaning that we still derive acceptable abundances for a majority of the spectra. Based on this investigation, the influence of S/N of the determined abundances of the high-S/N stars in this paper, is negligible, but it might be significant for some of the bulge spectra of Paper II having S/N<20.
	\item In Table \ref{tab:ges} in addition to the stellar parameters, also list the abundances of the three stars overlapping between \citet{2015A&A...582A..49H,2014A&A...564A.133J,2015A&A...582A..81J} and our sample. Our abundances fall within the uncertainties of the benchmark values with one obvious and one more subtle exception. Firstly, our derived magnesium abundance of $\mu$Leo is much lower, than the value from \citet{2015A&A...582A..81J}. We cannot find any reason for this, but note that the benchmark value would place $\mu$Leo significantly above the [Mg/Fe] vs. [Fe/H] trend if Figure \ref{fig:abundances} at ([Fe/H];[Mg/Fe])$=(0.20;0.31)$, while our result is more following the rest of the disk stars with ([Fe/H];[Mg/Fe])$=(0.20;0.03)$. Secondly, our magnesium abundance as derived from the FIES-spectrum of Arcturus is also slightly lower than the benchmark value including its uncertainty. The abundance as derived from the \citet{2000vnia.book.....H} atlas spectrum, however, is within the uncertainties of the benchmark value. The magnesium abundances as derived from the FIES-spectrum and the atlas-spectrum are deviating more than the abundances of oxygen, calcium, and titanium. Since the magnesium lines, as described in Section \ref{sec:linedata}, have a neighboring autoionizing line whose curving influence on the continuum is hard to get rid of, it is not unexpected for the magnesium abundance of these high-S/N spectra to show higher uncertainty than the other abundances. 
	\item The uncertainties of determined abundances are often dominated by the uncertainties of the stellar parameters, and as is suggested by the similar shapes of the parameter-boxes and the abundance-boxes in Figure \ref{fig:snr}, so is likely the case in this study. To estimate these uncertainties, often a table like Table \ref{tab:uncertainties} is used.	
	    \begin{table}[ht]
	    \caption{The uncertainties of the derived abundances of a typical star (our FIES Arcturus spectrum) for changes in the derived stellar parameters.}
	    \begin{tabular}{l r r r}
	    \hline
	    \hline
	    Uncertainty & $\delta$ A(O) & $\delta$ A(Mg) & $\delta$ A(Ti) \\
	    \hline
	    $\delta T_{\mathrm{eff}}=-50$~K &   $-0.01$ &    +0.01  &   $-0.07$ \\
	    $\delta T_{\mathrm{eff}}=+50$~K &    +0.01  & $\pm0.00$ &    +0.07  \\
	    $\delta \log g=-0.15$           &   $-0.07$ &   $-0.02$ &   $-0.01$ \\
	    $\delta \log g=+0.15$           &    +0.06  &    +0.03  &    +0.01  \\
	    $\delta$[Fe/H]$=-0.05$          &    +0.02  & $\pm0.00$ & $\pm0.00$ \\
	    $\delta$[Fe/H]$=+0.05$          &   $-0.01$ & $\pm0.00$ & $\pm0.00$ \\
	    $\delta v_{\mathrm{mic}}=-0.1$  & $\pm0.00$ &     0.01  &    +0.02  \\
	    $\delta v_{\mathrm{mic}}=+0.1$  &   $-0.01$ &   $-0.01$ &   $-0.02$ \\
	    \hline
	    \end{tabular}
	    \label{tab:uncertainties}
	    \end{table}
	    This table is best used as a way of telling which abundances are most uncertain, and which parameter is mainly influencing the different abundances. For example, our probable systematic error of +0.1 dex in determining the $\log g$ is mainly going to influence the oxygen abundance, while the titanium abundance is not affected by this. Estimating the total uncertainty in the abundance determination by adding the different uncertainties in quadrature would overestimate the uncertainties, since the parameters are coupled and deriving an incorrect temperature possibly would lead to an incorrect metallicity, for example. Furthermore, such an exercise would be based on the uncertainties of the stellar parameters which in themselves are hard to estimate (compare Section \ref{sec:random}).
\end{itemize}

To conclude we can estimate expected typical uncertainties of the determined abundances to be almost of the order of 0.1 dex for magnesium, and likely lower for the others.

\section{Results} \label{sec:results}
\begin{figure}[h]
\centering
\includegraphics[width=90mm]{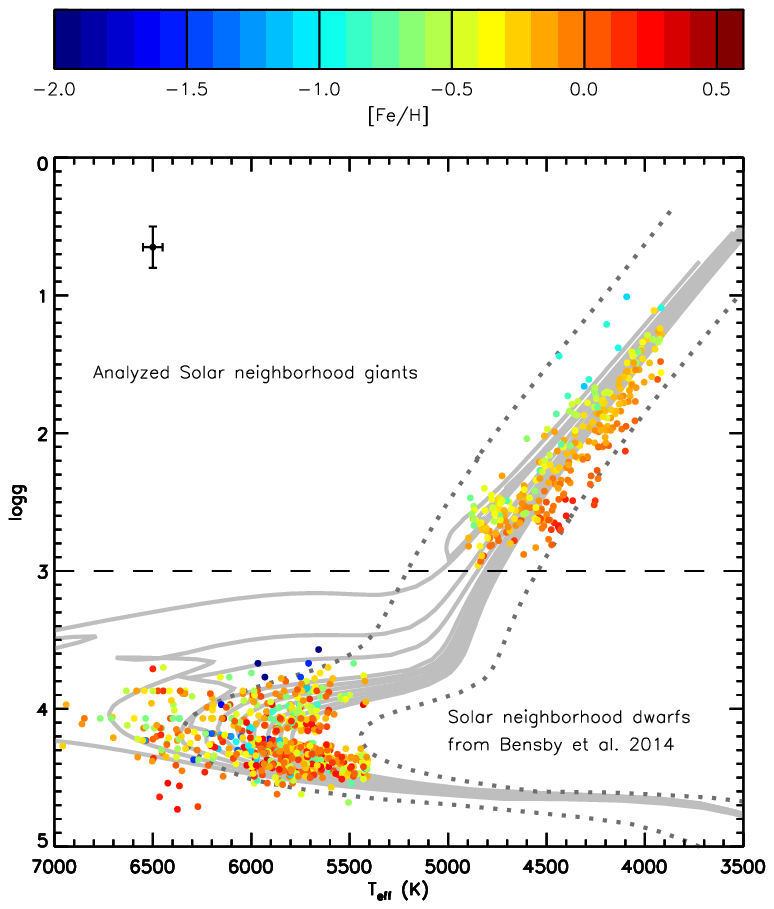}
\caption{HR-diagrams for the observed giants and the 604 Solar neighborhood dwarfs of \citet{2014A&A...562A..71B} with $T_{\mathrm{eff}}>5400$~K. As a guide for the eye, isochrones with [Fe/H]=0.0 and ages 1-10 Gyr are plotted using solid light gray lines. Furthermore, one isochrone with [Fe/H]=-1.0 and age 10 Gyr, and one with [Fe/H]=+0.5 and age 10 Gyr are plotted using dotted dark grey lines \citep{2012MNRAS.427..127B}. The parameters of the analyzed giants are determined from spectroscopy as described in the text. They line up as expected from isochrones in both temperature, and surface gravity, as well as metallicity. Expected, typical uncertanties are marked in the top left corner of the plot.}
\label{fig:hr-dia}
\end{figure}

The HR-diagram based on the spectroscopically determined stellar parameters of the observed giants are shown in Figure \ref{fig:hr-dia}. Also shown in the figure are the 604 solar neighborhood dwarf stars of \citet{2014A&A...562A..71B} with $T_{\mathrm{eff}}>5400$~K  and a collection of isochrones \citep{2012MNRAS.427..127B}. 

\begin{figure*}[htp]
\centering
\includegraphics[width=170mm]{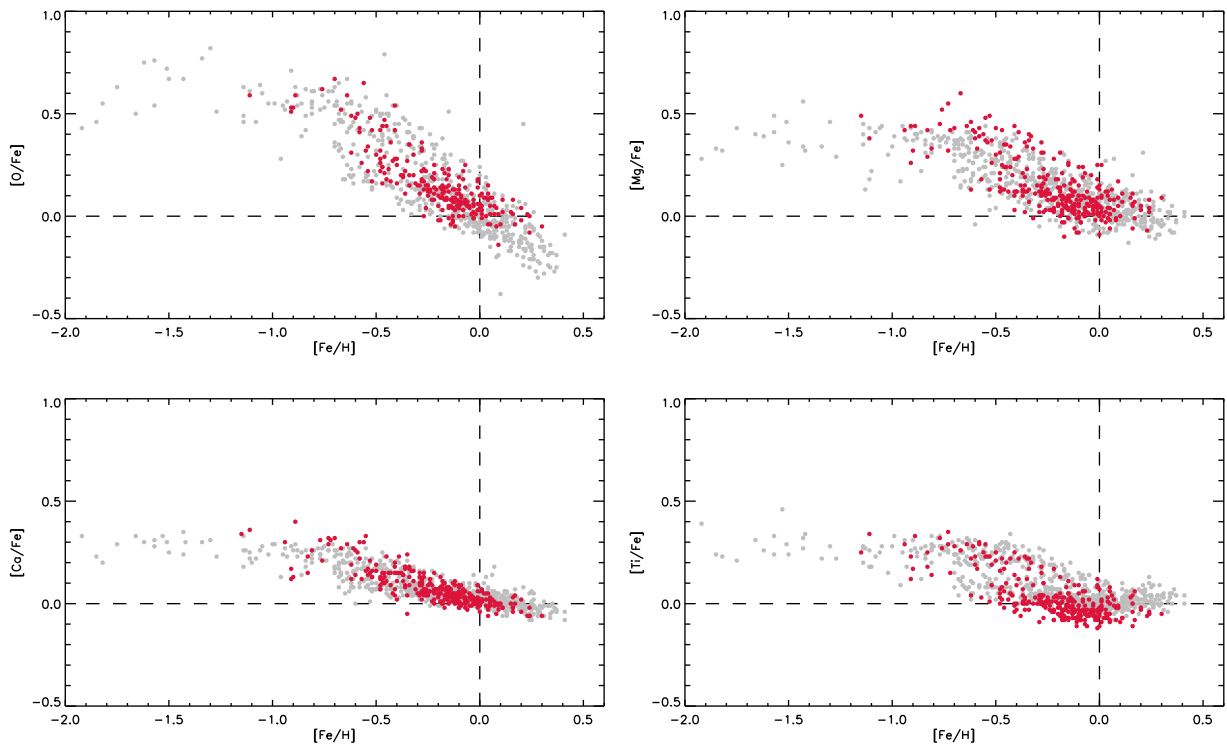}
\caption{[X/Fe] for the observed giants in red, compared to abundances of the 604 solar neighborhood dwarf stars of \citet{2014A&A...562A..71B} with $T_{\mathrm{eff}}>5400$~K in gray. Especially for O and Ti a clear separation of thin and thick disk type abundances can be seen in our sample. Since our sample is mostly made up of the very brightest, closest, giants, it is not surprising that a majority of the stars show thin disk type chemistry. In the plots we use A(O)$_{\odot}=8.69$, A(Mg)$_{\odot}=7.60$, A(Ca)$_{\odot}=6.34$, A(Ti)$_{\odot}=4.95$, and A(Fe)$_{\odot}=7.50$ \citep{2009ARA&A..47..481A}.}
\label{fig:abundances}
\end{figure*}

Our derived abundances for the giants can be seen in Table \ref{tab:abundances} (Online material) and in Figure \ref{fig:abundances}. As a comparison, the corresponding abundances as derived in a local disk sample of dwarf stars by \citet{2014A&A...562A..71B} is also shown in the figure.

\section{Discussion} \label{sec:discussion}
The aim of this analysis is twofold: to find as accurate stellar parameters and abundances as possible from giants, but most importantly, to collect a sample of homogeneously analyzed Solar neighborhood targets similar to the bulge targets that will be presented in Paper II. The second aim means that we have to restrict our present analysis to the wavelength coverage of the bulge spectra even if the spectra analyzed here have a wider coverage. In turn, this optional restriction might lead to a possibly lower precision of the abundances in the Solar neighborhood sample that what would have been possible to attain using the entire coverage available from FIES, NARVAL, and ESpaDOnS. 

The low deviations of the temperatures and gravities determined when compared to the more accurate angular diameter and asteroseismic measurements, together with the alignment in of the measurements along the red giant branch and and spread in [Fe/H] in the HR-diagram (Figure \ref{fig:hr-dia}), give us assurance that the method used likely is finding accurate values for the stellar parameters.

Estimating the formal uncertainties in abundance determinations is difficult: in our case where we have very high S/N-spectra and we are only using spectral lines with precise atomic data that are believed to be unblended, the main uncertainty of the abundances comes from the uncertainties of the stellar parameters. As elaborated on in Section \ref{sec:uncertainties}, the often used way of adding the dependencies of the abundances on the individual stellar parameters (Table \ref{tab:uncertainties}), gives an overestimation on the total abundance uncertainty. All in all, our results are showing very similar trends as the carefully analyzed dwarfs in \citet{2014A&A...562A..71B} in Figure \ref{fig:abundances}, and also have similarly tight trends, which hints at high precision in our derived abundances. The mean of the uncertainties quoted in  \citet{2014A&A...562A..71B} for the 604 stars with $T_{\mathrm{eff}}>5400$~K are $\delta$A(O)~$\sim0.14$, $\delta$A(Mg)~$\sim0.06$, $\delta$A(Ca)~$\sim0.06$, $\delta$A(Ti)~$\sim0.07$, and our uncertainties are thus likely at the same order. In the case of magnesium our scatter is higher, hinting at a larger uncertainty, but in the case of oxygen our trend looks tighter, hinting at lower mean uncertainty than that of \citet{2014A&A...562A..71B}. This is not unexpected, since \citeauthor{2014A&A...562A..71B} use the very NLTE-sensitive oxygen triplet around 7771-7775~\AA~\citep{2016MNRAS.455.3735A}, although with corrections, making their oxygen abundances less precise than their other abundances, a fact that is reflected in their higher quoted uncertainty for oxygen as mentioned above.

Based on the comparison between our abundance trends using K-giants and those of the carefully analyzed dwarfs of \citet{2014A&A...562A..71B} we would like to claim that giant stars are as good as abundance indicators as dwarfs. This is an important point, since this opens up the usefulness of the giants due to their brightness. However, as described in Section \ref{sec:introduction}, care must be shown to possible blending lines, continuum fitting, and the atomic and/or molecular data for the used spectral lines.

\section{Conclusions} \label{sec:conclusions}
The main purpose of this paper is to analyze a sample of local disk K-giants to be used as a comparison sample for the similar sample bulge giants to be analyzed in Paper II. To be sure the local sample can be strictly differentially compared to the bulge sample, where photometry is difficult/uncertain, we have chosen a strictly spectroscopic approach. Furthermore we have chosen to restrict our analysis to the useful part of the bulge spectra: 5800~\AA-6800~\AA. When we compare the determined effective temperatures to those derived from angular diameter measurements, we reproduce these with a systematic difference of +10~K and a standard deviation of 53~K. The spectroscopic gravities are reproducing the ones determined from asteroseismology with a systematic offset of +0.10~dex and a standard deviation of 0.12~dex. Regarding the abundance trends, our sample of local disk giants is closely following that of the Solar neighborhood dwarfs of \citet{2014A&A...562A..71B} with similar spread and no obvious systematic differences.

Ideally, our sample should have consisted of more stars with typical thick disk abundances to better represent both local disk stellar populations, and we plan to expand the sample accordingly in the near future.

\begin{acknowledgements}
Anders Thygesen is thanked for offering the reduced spectra of their stars online, and for kindly providing the observation dates/times to enable a scan for telluric contamination. John Telting is thanked for help with the FIES-data, and in particular archive-data. This research has been partly supported by the Lars Hierta Memorial Foundation, and the Royal Physiographic Society in Lund through Stiftelsen Walter Gyllenbergs fond and M\"arta och Erik Holmbergs donation. H.H., P.J., and N.R. acknowledges support from the Swedish Research Council, VR (project numbers 2011-4206, 2014-5640, and 2015-4842). This publication made use of the SIMBAD database, operated at CDS, Strasbourg, France, NASA's Astrophysics Data System, and the VALD database, operated at Uppsala University, the Institute of Astronomy RAS in Moscow, and the University of Vienna.
\end{acknowledgements}


\bibliographystyle{aa}

\onllongtab{
\begin{longtable}{l l l l r r r l}
\caption{Basic data for the observed giants. Coordinates and magnitudes are taken from the SIMBAD database, while the radial velocities are measured from the spectra. The S/N per datapoint is measured by the IDL-routine \texttt{der\textunderscore snr.pro}, see \href{http://www.stecf.org/software/ASTROsoft/DER\textunderscore SNR}{http://www.stecf.org/software/ASTROsoft/DER\textunderscore SNR}.}\\
\hline
\hline
HIP/KIC/TYC & Alternative name & RA (J2000) & Dec (J2000) & $V$ & $v_{\mathrm{rad}}$ & S/N &  Source\\
        &                  & (h:m:s)    & (d:am:as)   &     & km/s\\
\hline
\endfirsthead

\caption{Continued.}\\
\hline
HIP/KIC/TYC & Alternative name & RA (J2000) & Dec (J2000) & $V$ & $v_{\mathrm{rad}}$ & S/N & Source\\
        &                  & (h:m:s)    & (d:am:as)   &     & km/s\\
\hline
\endhead     
HIP1692 &           HD1690 & 00:21:13.32713 & -08:16:52.1625 &  9.18 &   18.37 &  114 & FIES-archive \\
HIP9884 &           alfAri & 02:07:10.40570 & +23:27:44.7032 &  2.01 &  -14.29 &   90 & PolarBase \\
HIP10085 &          HD13189 & 02:09:40.17260 & +32:18:59.1649 &  7.56 &   26.21 &  156 & FIES-archive \\
HIP12247 &            81Cet & 02:37:41.80105 & -03:23:46.2201 &  5.66 &    9.34 &  176 & FIES-archive \\
HIP28417 &          HD40460 & 06:00:06.03883 & +27:16:19.8614 &  6.62 &  100.64 &  121 & PolarBase \\
HIP33827 &           HR2581 & 07:01:21.41827 & +70:48:29.8674 &  5.69 &  -17.99 &   79 & PolarBase \\
HIP35759 &          HD57470 & 07:22:33.85798 & +29:49:27.6626 &  7.67 &  -30.19 &   85 & PolarBase \\
HIP37447 &           alfMon & 07:41:14.83257 & -09:33:04.0711 &  3.93 &   11.83 &   71 & Thygesen et al. (2012)\\
\end{longtable}
}

\onltab{
\begin{table*}
\caption{Atomic data for the spectral lines used in the analysis. The first part list the lines used in the determination of the stellar parameters and calcium abundance, while the second part list the lines used to determine the oxygen, magnesium, and titanium abundances. All atomic data are collected by the Gaia-ESO line list group \citep{2015PhyS...90e4010H}. For the three \ion{Ca}{i}-lines marked with asterisks only the gravity-sensitive wings are used. The references are for wavelength, $\log gf$, and excitation energy respectively. In cases where several references are given for a quantity the value listed is a mean of the reference values.}
\begin{tabular}{c c c c c }
\hline
\hline
Element & Wavelength & $\log gf$ & $\chi_{\mathrm{exc}}$ & References\\
        & (Å) (air)  &            & (eV)                  &      \\
\hline
\ion{Fe}{i} &   5778.4533 &    -3.430 &     2.588 & 1, 2, 1\\
\ion{Fe}{i} &   5855.0758 &    -1.478 &     4.608 & 1, 2, 1\\
\ion{Fe}{i} &   6012.2098 &    -4.038 &     2.223 & 1, 3, 1\\
\ion{Fe}{i} &   6027.0508 &    -1.089 &     4.076 & 1, 4, 1\\
\ion{Fe}{i} &   6120.2464 &    -5.970 &     0.915 & 1, 5, 1\\
\ion{Fe}{i} &   6136.9938 &    -2.950 &     2.198 & 1, 4+6, 1\\
\ion{Fe}{i} &   6151.6173 &    -3.295 &     2.176 & 1, 3+4+6, 1\\
\ion{Fe}{i} &   6165.3598 &    -1.473 &     4.143 & 1, 4, 1\\
\ion{Fe}{i} &   6173.3343 &    -2.880 &     2.223 & 1, 6, 1\\
\ion{Fe}{i} &   6213.4294 &    -2.481 &     2.223 & 1, 4, 1\\
\ion{Fe}{i} &   6271.2779 &    -2.703 &     3.332 & 1, 2, 1\\
\ion{Fe}{i} &   6322.6850 &    -2.430 &     2.588 & 1, 4+7, 1\\
\ion{Fe}{i} &   6335.3299 &    -2.177 &     2.198 & 1, 4, 1\\
\ion{Fe}{i} &   6411.6480 &    -0.596 &     3.654 & 1, 3+4+8, 1\\
\ion{Fe}{i} &   6518.3657 &    -2.438 &     2.832 & 1, 2+4, 1\\
\ion{Fe}{i} &   6581.2092 &    -4.679 &     1.485 & 1, 3, 1\\
\ion{Fe}{i} &   6593.8695 &    -2.420 &     2.433 & 1, 4+6, 1\\
\ion{Fe}{i} &   6609.1097 &    -2.691 &     2.559 & 1, 4+7, 1\\
\ion{Fe}{i} &   6633.7487 &    -0.799 &     4.559 & 1, 4, 1\\
\ion{Fe}{i} &   6739.5204 &    -4.794 &     1.557 & 1, 3, 1\\
\ion{Fe}{i} &   6793.2582 &    -2.326 &     4.076 & 1, 2, 1\\
\ion{Fe}{i} &   6810.2622 &    -0.986 &     4.607 & 1, 4, 1\\
\ion{Fe}{i} &   6828.5912 &    -0.820 &     4.638 & 9, 10, 1\\
\ion{Fe}{i} &   6837.0056 &    -1.687 &     4.593 & 1, 2, 1\\
\ion{Fe}{i} &   6843.6554 &    -0.730 &     4.549 & 1, 11, 1\\
\ion{Fe}{ii} &  5991.3721 &    -3.540 &     3.153 & 12, 13, 14\\
\ion{Fe}{ii} &  6084.1030 &    -3.790 &     3.200 & 12, 13, 14\\
\ion{Fe}{ii} &  6113.3192 &    -4.140 &     3.221 & 12, 13, 14\\
\ion{Fe}{ii} &  6149.2459 &    -2.690 &     3.889 & 12, 13, 14\\
\ion{Fe}{ii} &  6247.5590 &    -2.300 &     3.892 & 12, 13, 14\\
\ion{Fe}{ii} &  6432.6772 &    -3.570 &     2.891 & 12, 13, 14\\
\ion{Fe}{ii} &  6456.3805 &    -2.050 &     3.903 & 12, 13, 14\\
\ion{Ca}{i} &   5867.5620 &    -1.570 &     2.933 & 15, 15, 15\\
\ion{Ca}{i}$^*$ &   6122.2170 &    -0.380 &     1.886 & 16, 17, 16\\
\ion{Ca}{i}$^*$ &   6162.1730 &    -0.170 &     1.899 & 16, 17, 16\\
\ion{Ca}{i} &   6166.4390 &    -1.142 &     2.521 & 18+19, 18, 18+19\\
\ion{Ca}{i} &   6169.0420 &    -0.797 &     2.523 & 18+19, 18, 18+19\\
\ion{Ca}{i}$^*$ &   6439.0750 &     0.390 &     2.526 & 18+19, 18, 18+19\\
\ion{Ca}{i} &   6499.6500 &    -0.818 &     2.523 & 18+19, 18, 18+19\\
\hline
\ion{O}{i}  &   6300.3038 &   -9.715 &   0.000 & 20, 21+22 , 23\\
\ion{Mg}{i} &   6318.7170 &   -2.020 &   5.108 & 23, 24    , 23\\
\ion{Mg}{i} &   6319.2370 &   -2.242 &   5.108 & 23, 24    , 23\\
\ion{Mg}{i} &   6319.4930 &   -2.719 &   5.108 & 23, 24    , 23\\
\ion{Ti}{i} &   6091.1710 &   -0.320 &   2.267 & 25, 26    , 25\\
\ion{Ti}{i} &   6092.7918 &   -1.380 &   1.887 & 25, 26    , 25\\
\ion{Ti}{i} &   6336.0993 &   -1.690 &   1.443 & 25, 26    , 25\\
\ion{Ti}{i} &   6395.4718 &   -2.540 &   1.503 & 25, 26    , 25\\
\hline
\label{tab:linedata}
\end{tabular}
\tablebib{\\
(1) \citet{K07};
(2) \citet{BK};
(3) \citet{BKK};
(4) \citet{BWL};
(5) \citet{GESB86};
(6) \citet{GESB82c};
(7) \citet{GESB82d};
(8) \citet{GESHRL14};
(9) \citet{FMW};
(10) \citet{MRW};
(11) \citet{2014MNRAS.441.3127R};
(12) \citet{2013ApJS..204....1N};
(13) \citet{2009A&A...497..611M};
(14) \citet{K13};
(15) \citet{S};
(16) \citet{SN};
(17) \citet{2009AA...502..989A};
(18) \citet{SR};
(19) \citet{Sm};
(20) \citet{WSG};
(21) \citet{2000MNRAS.312..813S};
(22) \citet{GESMCHF};
(23) \citet{NIST10};
(24) Pehlivan et al. (in prep);
(25) \citet{LGWSC};
(26) \citet{2013ApJS..205...11L}
}
\end{table*}
}

\onllongtab{
\begin{longtable}{l c c c c c c c c c c }
\caption{Determined stellar parameters and abundances for the observed giants. The reference values for $T_{\mathrm{eff}}$ are from \citet{2003AJ....126.2502M} and the reference values for $\log g$ are from \citet{2014ApJS..211....2H}}\\
\hline
\hline
HIP/KIC/TYC & $T_{\mathrm{eff, ref}}$ & $T_{\mathrm{eff}}$ & $\log g_{\mathrm{ref}}$ & $\log g$ & [Fe/H] & $v_{\mathrm{mic}}$ & A(O) & A(Mg) & A(Ca) & A(Ti)\\ 
\hline
\endfirsthead

\caption{Continued.}\\
\hline
HIP/KIC/TYC & $T_{\mathrm{eff, ref}}$ & $T_{\mathrm{eff}}$ & $\log g_{\mathrm{ref}}$ & $\log g$ & [Fe/H] & $v_{\mathrm{mic}}$ & A(O) & A(Mg) & A(Ca) & A(Ti)\\ 
\hline
\endhead     
             HIP1692 &           ... &   4216 &                   ... &   1.79 &  -0.29 &   1.55 &    ... &   7.50 &   6.10 &   4.68 \\
             HIP9884 & 4493 $\pm$ 55 &   4464 &                   ... &   2.27 &  -0.24 &   1.34 &   8.57 &   7.47 &   6.16 &   4.72 \\
            HIP10085 &           ... &   4062 &                   ... &   1.44 &  -0.35 &   1.63 &    ... &   7.37 &   6.03 &   4.60 \\
            HIP12247 &           ... &   4790 &                   ... &   2.71 &  -0.07 &   1.40 &   8.69 &   7.53 &   6.33 &   4.83 \\
            HIP28417 &           ... &   4746 &                   ... &   2.56 &  -0.28 &   1.40 &   8.63 &   7.44 &   6.16 &   4.67 \\
            HIP33827 &           ... &   4235 &                   ... &   1.99 &  -0.02 &   1.50 &    ... &   7.65 &   6.34 &   4.87 \\
            HIP35759 &           ... &   4606 &                   ... &   2.47 &  -0.18 &   1.42 &   8.74 &   7.63 &   6.25 &   4.80 \\
            HIP37447 &           ... &   4758 &                   ... &   2.73 &  -0.07 &   1.35 &   8.68 &   7.55 &   6.28 &   4.78 \\
\end{longtable}
}

\end{document}